\documentclass[a4paper,11pt]{article}
\usepackage{pos}
\allowdisplaybreaks

\title{Dark matter phenomenology in Z’2 broken singlet extended 2HDM}
\ShortTitle{Dark matter phenomenology in a 2HDMS}

\author*[a]{Julia Anabell Ziegler}
\author[b]{Juhi Dutta}  
\author[a]{Jayita Lahiri}
\author[c,d]{Cheng Li}
\author[a,c]{Gudrid Moortgat-Pick}
\author[a]{Sheikh Farah Tabira}
\affiliation[a]{II. Institut f{\"u}r Theoretische Physik,  Universit{\"a}t Hamburg, \\
    Luruper Chaussee 149, 22761 Hamburg, Germany}
\affiliation[b]{Homer L. Dodge Department of Physics and Astronomy, University of Oklahoma, \\
    Norman, OK 73019, USA}
\affiliation[c]{Deutsches Elektronen-Synchrotron DESY, \\
    Notkestr. 85, 22607 Hamburg, Germany}
\affiliation[d]{School of Science, Sun Yat-Sen University, \\
    Gongchang Road 66, 518107 Shenzhen, China}
\emailAdd{julia.ziegler@desy.de}
\emailAdd{juhi.dutta@ou.edu}
\emailAdd{jayita.lahiri@desy.de}
\emailAdd{gudrid.moortgat-pick@desy.de}
\emailAdd{cheng.li@desy.de}
\emailAdd{sheikh.farah.tabira@desy.de}

\abstract{Many different approaches have been made to explain the nature of dark matter (DM), but it remains and unsolved mystery of our universe. In this work we examine a type II two-Higgs-doublet model extended by a complex singlet (2HDMS), where the pseudo-scalar component of the singlet acts as a natural DM candidate. The DM candidate is stabilized by a $Z'_2$ symmetry, which is broken spontaneously by the singlet acquiring a vacuum expectation value (vev). This vev in turn causes the scalar component of the singlet to mix with the scalar components of the two doublets, which results in three scalar Higgs particles. Additionally we aim to include an excess around $95 \, \text{GeV}$, which was observed at CSM and LEP and can be explained by one of the three scalar Higgs particles. \\ 
After introducing the model, we apply experimental and theoretical constraints and find a viable benchmark point. We then look into the DM phenomenology as well as collider phenomenology.}

\FullConference{The European Physical Society Conference on High Energy Physics (EPS-HEP2023)\\
 21-25 August 2023\\
Hamburg, Germany\\}

\renewcommand{\hookAfterAbstract}{%
\par\bigskip
\text{DESY-23-167}\\
}

\renewcommand{\logo}{\relax}

\begin{document}
\maketitle
\section{Introduction}
Even though the Standard Model (SM) of particle physics has shown huge success in providing experimental predictions, there are many phenomena it can not explain, such as dark matter (DM), the matter-antimatter asymmetry, neutrino masses and other problems. This has led to many theories beyond the SM (BSM). 

One natural choice for BSM models is the two-Higgs-doublet model (2HDM), which instead of one Higgs doublet, as in the SM, contains two Higgs doublets, leading to a richer phenomenology. A good overview can be found in ref.~\cite{Branco}. This model can be further extended with a complex singlet, leading to a still richer particle content in the Higgs sector and a natural DM candidate. This model was studied in ref.~\cite{Baum_Shah}.

The aim of this work is to not only explain DM but also an excess around $95 \, \text{GeV}$ observed at the Compact Muon Solenoid (CMS)~\cite{96_excess_CMS} and at the Large Electron–Positron Collider (LEP)~\cite{96_excess_LEP}. Therefore we consider a 2HDMS, where the pseudo-scalar component of the singlet acts as a natural DM candidate, since it is massive, electrically neutral, colorless and stable. The stabilization of the DM is achieved through a $Z'_2$ symmetry, which is spontaneously broken by the vacuum expectation value (vev) of the singlet. This vev leads to a mixing of the scalar sector. The total particle content of the Higgs sector results in three scalar, two pseudo-scalar, and two charged Higgs particles. Of the three scalars, the lightest one is chosen to have a mass of $95 \, \text{GeV}$ in accordance with the excess explained above, the second lightest is chosen to be SM-like with a mass of $125 \, \text{GeV}$ and the heavy one is chosen to have a mass of $900 \, \text{GeV}$. 

The possibility to accommodate the excess around $95 \, \text{GeV}$ in a 2HDMS was investigated in ref.~\cite{Steven_Cheng_Li}. The DM phenomenology in a 2HDMS without the singlet obtaining a vev was investigated in ref.~\cite{merle_juhi_2HDMS}. 

This work wraps up the study done in ref.~\cite{dutta2023dark}. We first introduce the model and the considered constraints. Then we look into the DM phenomenology, namely relic density, indirect detection and direct detection of DM. Finally we look into the collider phenomenology at future electron and muon colliders and at the High Luminosity Large Hadron Collider (HL-LHC). We then conclude our work.

\section{The 2HDMS model}
As mentioned above, we consider a type II 2HDMS, where the singlet acquires a vev. The 2HDM part of the potential is symmetric under a $U(1)$ symmetry, to avoid charge-parity (CP) violation, and under a $Z_2$ symmetry, to avoid flavor-changing neutral currents (FCNC). The singlet part of the potential is symmetric under a $Z'_2$ symmetry, to stabilize the DM candidate. The full Higgs sector potential is the sum of the 2HDM and the singlet potential and can be written as
\begin{subequations}
\begin{align}\label{eq:2HDMS_potentail}
    V &= V_{2HDM} + V_{S} \\
    \label{eq:2HDM_potential}V_{2HDM} &= m_{11}^2 \Phi_1^{\dagger} \Phi_1 + m_{22}^2 \Phi_2^{\dagger} \Phi_2 - [m_{12}^2 \Phi_1^{\dagger} \Phi_2 + h.c. ] + \frac{\lambda_1}{2} (\Phi_1^{\dagger} \Phi_1)^2 \nonumber \\ 
    & \quad + \frac{\lambda_2}{2}(\Phi_2^{\dagger} \Phi_2)^2  + \lambda_3 (\Phi_1^{\dagger} \Phi_1) (\Phi_2^{\dagger} \Phi_2) + \lambda_4 (\Phi_1^{\dagger} \Phi_2) (\Phi_2^{\dagger} \Phi_1) \nonumber \\
    &\quad + \left[ \frac{\lambda_5}{2} (\Phi_1^{\dagger} \Phi_2)^2 + h.c. \right]  \\
    V_{S} &= m_S^2 S^{\dagger} S + \left[ \frac{m_S'^2}{2} S^2 + h.c. \right] \nonumber \\ 
    & \quad + \left[ \frac{\lambda_1''}{24} S^4 + h.c. \right] + \left[ \frac{\lambda_2''}{6} (S^2 S^{\dagger} S) + h.c. \right] + \frac{\lambda_3''}{4}(S^{\dagger} S)^2 \nonumber \\
    & \quad + S^{\dagger} S [\lambda_1' \Phi_1^{\dagger} \Phi_1 + \lambda_2' \Phi_2^{\dagger} \Phi_2] + [S^2(\lambda_4' \Phi_1^{\dagger} \Phi_1 + \lambda_5' \Phi_2^{\dagger} \Phi_2) + h.c.] , 
\end{align}
\end{subequations}
where $h.c.$ stands for the hermitian conjugate, $\Phi_{1,2}$ denote the two doublets and $S$ denotes the singlet. For simplicity we set the parameters $\lambda_1''=\lambda_2''$.

After spontaneous symmetry breaking both doublets and the singlet obtain vevs. They can be expanded around the vevs and written in terms of real and imaginary component. Where the real components give rise to the scalar particles and the imaginary components give rise to the pseudo-scalar particles. The charged particles results from the upper components of the doublets and do no acquire a vev. The doublet and singlet fields can then be written as
\begin{subequations}
\begin{align}
    \Phi_i &= \begin{pmatrix} \phi_i^+ \\ \frac{1}{\sqrt{2}}(v_i + \rho_i + i \eta_i) \end{pmatrix} & \langle \Phi_i \rangle &= \begin{pmatrix} 0 \\ \frac{v_i}{\sqrt{2}} \end{pmatrix} , \quad i=1,2 \\
    S &= \frac{1}{\sqrt{2}}(v_S + \rho_S + i A_S) & \langle S \rangle &= \frac{v_S}{\sqrt{2}} ,
\end{align}
\end{subequations}
where $v_{1,2}$ denote the vevs of the two doublets and $v_S$ the singlet vev. The imaginary component of the singlet $A_S$ is the DM candidate.

After diagonalization of the mass matrix we are left with three scalars $h_1$, $h_2$, $h_3$, two pseudo-scalars $A$, $A_S$, and two charged Higgs particles $H^\pm$, as well as two charged Goldstone bosons $G^\pm$ and a pseudo-scalar Goldstone boson $G^0$. The mixing of these mass eigenstates is as follows, further details can be found in ref.~\cite{dutta2023dark}:
\begin{align}
    \begin{pmatrix} h_1 \\ h_2 \\ h_3 \end{pmatrix}= R \begin{pmatrix}
    \rho_1 \\ \rho_2 \\ \rho_S
    \end{pmatrix}, \quad
    \begin{pmatrix} A \\ G^0 \end{pmatrix}= R^A \begin{pmatrix}
    \eta_1 \\ \eta_2 \\
    \end{pmatrix}, \quad
    \begin{pmatrix} A_S \end{pmatrix}=\begin{pmatrix} A_S \end{pmatrix}, \quad
    \begin{pmatrix} H^\pm \\ G^\pm \end{pmatrix}= R^\pm \begin{pmatrix}
    \phi_1^+ \\ \phi_2^+ 
    \end{pmatrix},    
\end{align}
where $R$ is the scalar, $R^A$ the pseudo-scalar and $R^\pm$ the charged mixing matrix and $A_S$ does not mix with the other mass eigenstates.

In a type II 2HDMS the up-type quarks couple to the second doublet $\Phi_2$ and the down-type quarks and leptons couple to the first doublet $\Phi_1$ to obtain their masses. The singlet has no direct coupling to the SM particles. Hence the pseudo-scalar DM candidate can couple to these only via the exchange of one of the scalars $h_1$, $h_2$, $h_3$.

\subsection{Benchmark point}
The free parameters in the interaction basis are:
\begin{align}
    \lambda_1, \lambda_2,  \lambda_3,  \lambda_4,  \lambda_5, m^2_{12}, \tan \beta, v_S,
    m^{2\prime}_S, \lambda^{\prime}_1, \lambda^{\prime}_2, \lambda^{\prime}_4, \lambda^{\prime}_5, \lambda^{\prime\prime}_1 = \lambda^{\prime\prime}_2, \lambda^{\prime\prime}_3.    
\end{align}
After a basis change these parameters can be expressed via the mass basis parameters:
\begin{align}
     m_{h_1}, m_{h_2}, &m_{h_3}, m_A, m_{A_S}, m_{H^\pm}, \delta_{14}'=\lambda^{\prime}_4 - \lambda^{\prime}_1, \delta_{25}'=\lambda^{\prime}_5 - \lambda^{\prime}_2, \nonumber \\
        &\tan \beta, v_S, c_{h_1 bb}, c_{h_1 tt}, \Tilde{\mu}^2, m_{S}'^2, alignm,   
\end{align}
where $m_X$ denotes the mass eigenvalue of the mass eigenstates $X=h_1,h_2,h_3,A,A_S,H^\pm$, the parameters $\delta_{14}'=\lambda^{\prime}_4 - \lambda^{\prime}_1$ and $\delta_{25}'=\lambda^{\prime}_5 - \lambda^{\prime}_2$, the reduced couplings $c_{h_1 dd}=\frac{R_{11}}{\cos\beta}$ and $c_{h_1 uu} = \frac{R_{12}}{\sin\beta}$, $\Tilde{\mu}^2=\frac{m^2_{12}}{\sin\beta\cos\beta}$, with $\tan(\beta)=\frac{v_2}{v_1}$, and the parameter assuring alignment limit $alignm=|\sin(\beta-\alpha_1-\alpha_3\cdot\text{sgn}(\alpha_2))| \approx 1$, with $\alpha_{1,2,3}$ being the angles of the scalar rotation matrix $R$. The motivation to choose these parameters as input parameters, as well as the full basis change equations can be found in ref.~\cite{dutta2023dark}.

Our analysis starts from the benchmark point BP1. The corresponding values of the mass basis parameters can be found in table \ref{tab:bp1}.
\begin{table}[h!]
    \centering
    \addtolength{\tabcolsep}{-3pt}
    \small
    \begin{tabular}{|c|c|c|c|c|}
        \hline
        $m_{h_1}$ & $m_{h_2}$ & $m_{h_3}$ & $m_A$ & $m_{A_S}$ \\ 
        \hline
        $95 \, \text{GeV}$ & $125.09 \, \text{GeV}$ & $900 \, \text{GeV}$ & $900 \, \text{GeV}$ & $325.86 \, \text{GeV}$ \\
        \hline
        $m_{H^\pm}$ & $m_{S}'^2$ & $\delta_{14}'$ & $\delta_{25}'$ &  $\tan(\beta)$\\
        \hline
        $900 \, \text{GeV}$ & $ -4.809 \times 10^4 \, \text{GeV}^2$ & $-9.6958$ & $0.2475$ &  $10$\\ 
        \hline  
        $v_S$ & $c_{h_1 bb}$ & $c_{h_1 tt}$ & $alignm$ & $\Tilde{\mu}^2$  \\
        \hline
        $ 239.86 \, \text{GeV}$ &  0.2096  & 0.4192  & $0.9998$ & $ 8.128 \times 10^5 \, \text{GeV}^2$  \\
        \hline 
    \end{tabular}
    \caption{Benchmark point BP1 in the mass basis}
    \label{tab:bp1}
\end{table}
This benchmark point was checked against theoretical constraints, such as bounded from below (bfb), unitarity (checked with \texttt{SPheno-v4.0.5}~\cite{Porod:2003um}) and vacuum stability constraints (checked with \texttt{EVADE}~\cite{Hollik:2018wrr,Ferreira:2019iqb}), as well as experimental constraints, such as constraints on the Higgs sector (checked with \texttt{HiggsTools}~\cite{Bahl:2022igd,Bechtle_2010,Bechtle:2013wla,Bechtle:2020pkv,Bechtle:2013xfa,Bechtle:2020uwn}), on the DM relic density (upper bounds from \texttt{Planck}~\cite{Aghanim:2018eyx}), on the DM indirect detection cross section (upper bounds from \texttt{Fermi-LAT}~\cite{Fermi-LAT:2011vow,Fermi-LAT:2016uux}) and on the DM direct detection cross section (upper bounds from \texttt{LUX-ZEPLIN}~\cite{LZ:2022ufs}). Furthermore the lightest scalar $h_1$ was chosen to have a mass of $95\,\text{GeV}$, in accordance with the excess at CMS and LEP, mentioned in the introduction. The second lightest scalar $h_2$ was chosen to be the SM-like Higgs with a mass of around $125 \,\text{GeV}$.

The programs used to produce the following results are \texttt{SARAH-v4.14.3}~\cite{Staub:2013tta} for the implementation of the model, \texttt{SPheno-v4.0.5}~\cite{Porod:2003um} for the generation of the spectrum, which in turn is used as input for \texttt{micrOmegas-v5.2.13}~\cite{mco2}, which is used for the DM phenomenology. For the collider phenomenology we use \texttt{MG5$\_$aMC$\_$v3.4.1}~\cite{Alwall:2014hca,Alwall:2011uj}, \texttt{Pythia$\_$v8}~\cite{Pythia8}, \texttt{Delphes-v3.5.0}~\cite{Selvaggi:2014mya}, \texttt{MadAnalysis-v5}~\cite{Conte:2012fm} and \texttt{WHIZARD}~\cite{Kilian:2007gr}.

\section{Results}
\subsection{Dark matter phenomenology}
\begin{figure}[ht]
    \centering
    \includegraphics[scale=0.45]{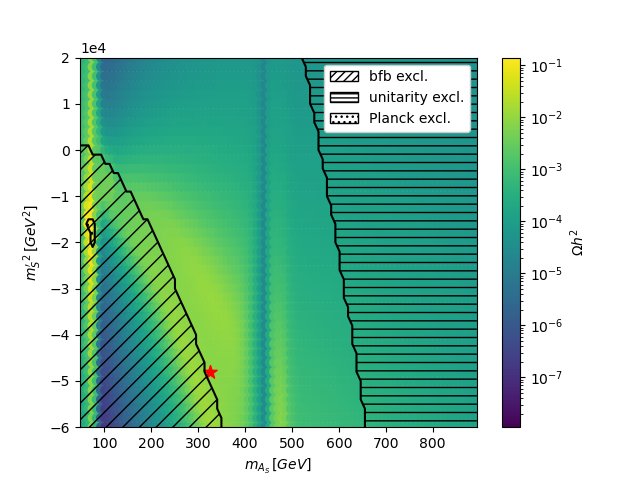}
    \includegraphics[scale=0.45]{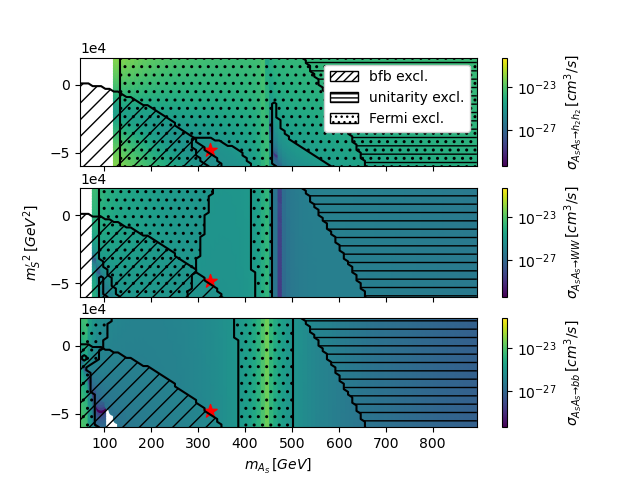}\\
    \includegraphics[scale=0.45]{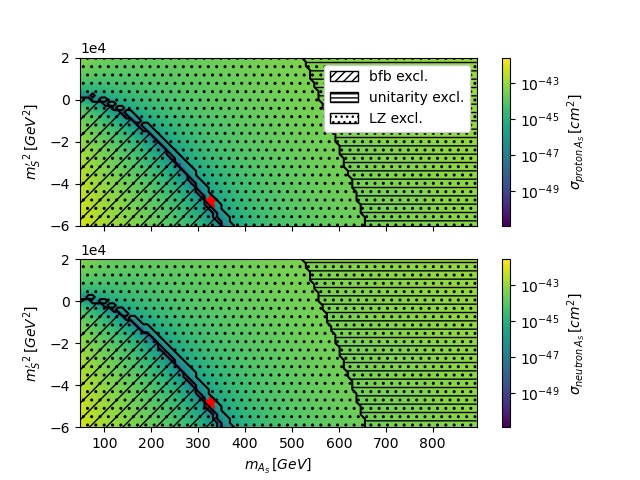}
    \includegraphics[scale=0.45]{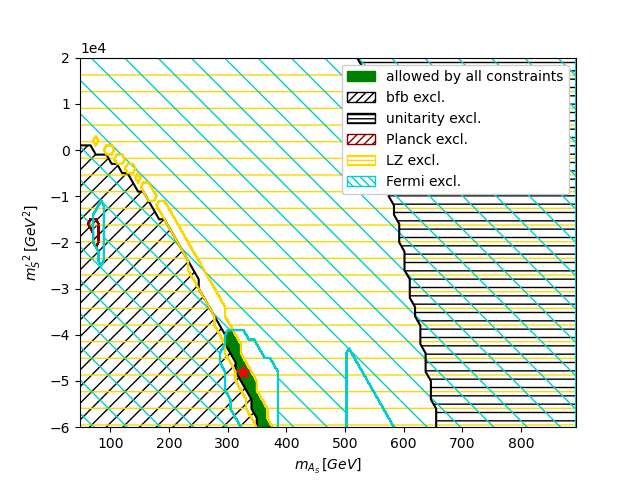}
    \caption{DM observables for varying the DM mass $m_{A_S}$ and $m_S'^2$ around BP1 (marked with a red star). The color coding shows the relic density $\Omega h^2$ (top left), indirect detection cross section $\sigma_{A_S A_S \rightarrow h_2h_2/WW/bb}$ (top right), direct detection cross section $\sigma_{\text{proton/neutron}A_S}$ (bottom left) and the allowed parameter regions under combination of all constraints (bottom right).}
    \label{fig:infl_mAS_mSp2}
\end{figure}

In this section we look into the change of DM observables, namely relic density, indirect detection cross section and direct detection cross section, under variation of the DM mass $m_{A_S}$ and the parameters $m_S'^2$ around the benchmark point BP1. The results can be seen in figure \ref{fig:infl_mAS_mSp2}.
The hatched areas show which regions are excluded by the theoretical and experimental constraints. The bfb, unitarity and vacuum stability constraints, as well as bounds on the Higgs sector apply to all three DM observables. Only those constraints which actually constrain the scanned parameter space are shown in the plots. In addition for the relic density (top left) constraints from \texttt{Planck}, for the indirect detection cross section (top right) constraints from \texttt{Fermi-LAT} and for the direct detection cross section (bottom left) constraints from \texttt{LUX-ZEPLIN} (\texttt{LZ}) are shown as dotted areas. In the plot on the bottom right, all these constraints are combined to show the allowed parameter space. As can be seen, under all constraints only a small strip around BP1 is allowed. The strongest constraints, on the parameter space shown here, come from \texttt{Fermi-LAT} and \texttt{LZ}. 

The behaviour of the relic density shows some interesting features. For example a dip around $m_{A_S}\approx 62.5\, \text{GeV}\approx\frac{m_{h_2}}{2}$, where resonant annihilation of two $A_S$ into one $h_2$ is possible. Between this peak and $m_{A_S}\approx 95\,\text{GeV}\approx m_{h_1}$ the relic density is quite high, then it drops again as the annihilation channel of two $A_S$ into two $h_1$ opens up and keeps the relic density at a low level, however increasing slowly for higher values of $m_{A_S}$. At $m_{A_S}\approx 450\, \text{GeV}\approx\frac{m_{h_3}}{2}$ another dip appears, as resonant annihilation of two $A_S$ into one $h_3$ decreases the relic density.

For the indirect detection cross section the three main annihilation channels $A_S A_S \rightarrow h_2h_2/ $ $ WW/ $ $ bb$ (from top to bottom) are shown. Some regions are white as the cross section is too low and no values are returned. The dip in the the relic density plot around $m_{A_S}\approx 450\, \text{GeV}\approx\frac{m_{h_3}}{2}$ shows up as a peak in the indirect detection plot, as both observables behave roughly inversely. (When more DM annihilates, the indirect detection cross section, which is just the annihilation cross section, grows. This also means that after annihilation less DM is left in the universe, hence relic density is decreased.) 

The direct detection cross section is shown for scattering of a DM particle on a proton and on a neutron. Both plots look very similar and show that BP1 lies right in a minimum of the cross section.

\subsection{Collider phenomenology}
\subsubsection{Future lepton colliders}
\begin{figure}[ht]
   \centering
   \includegraphics[scale=0.45]{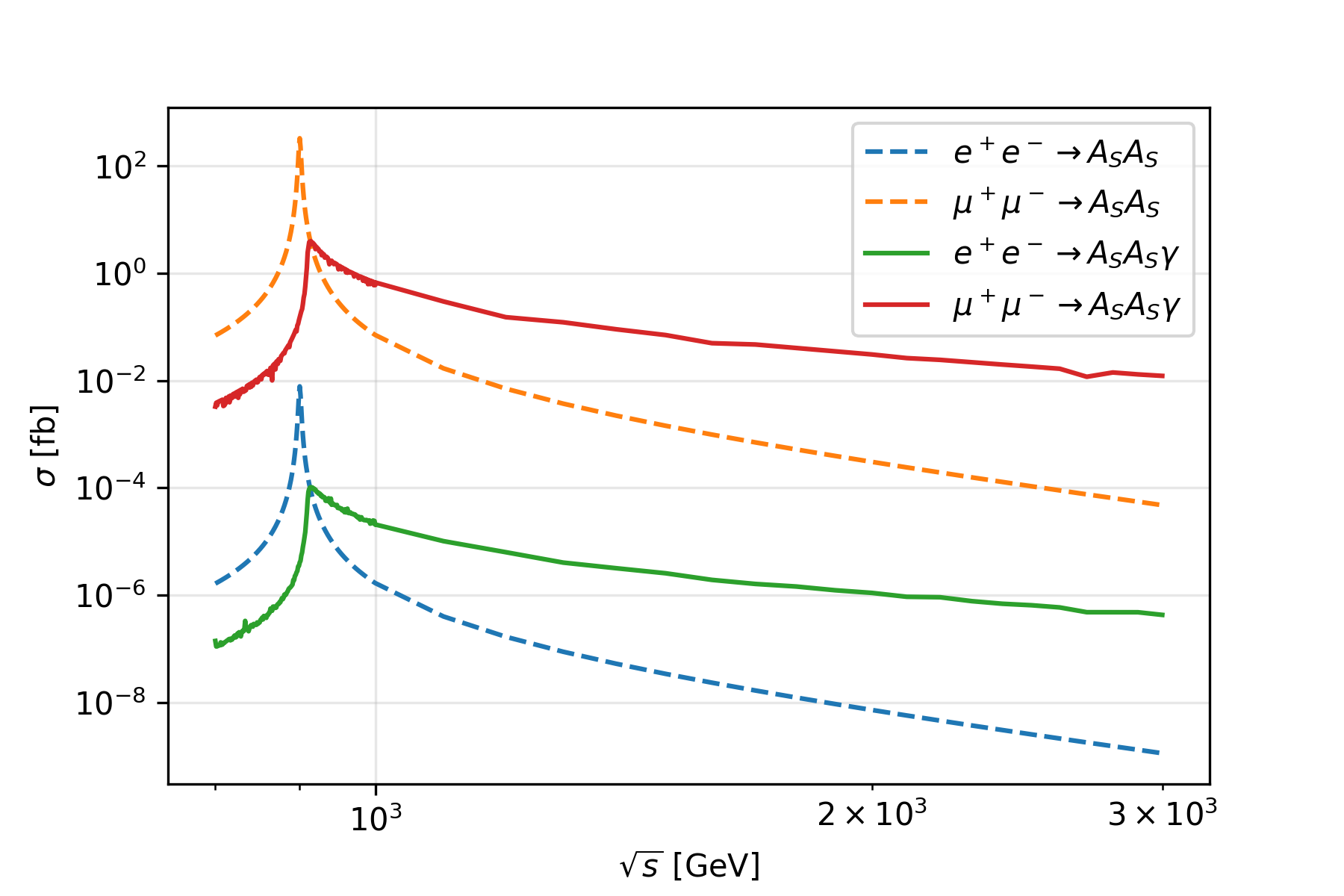}
   \includegraphics[scale=0.45]{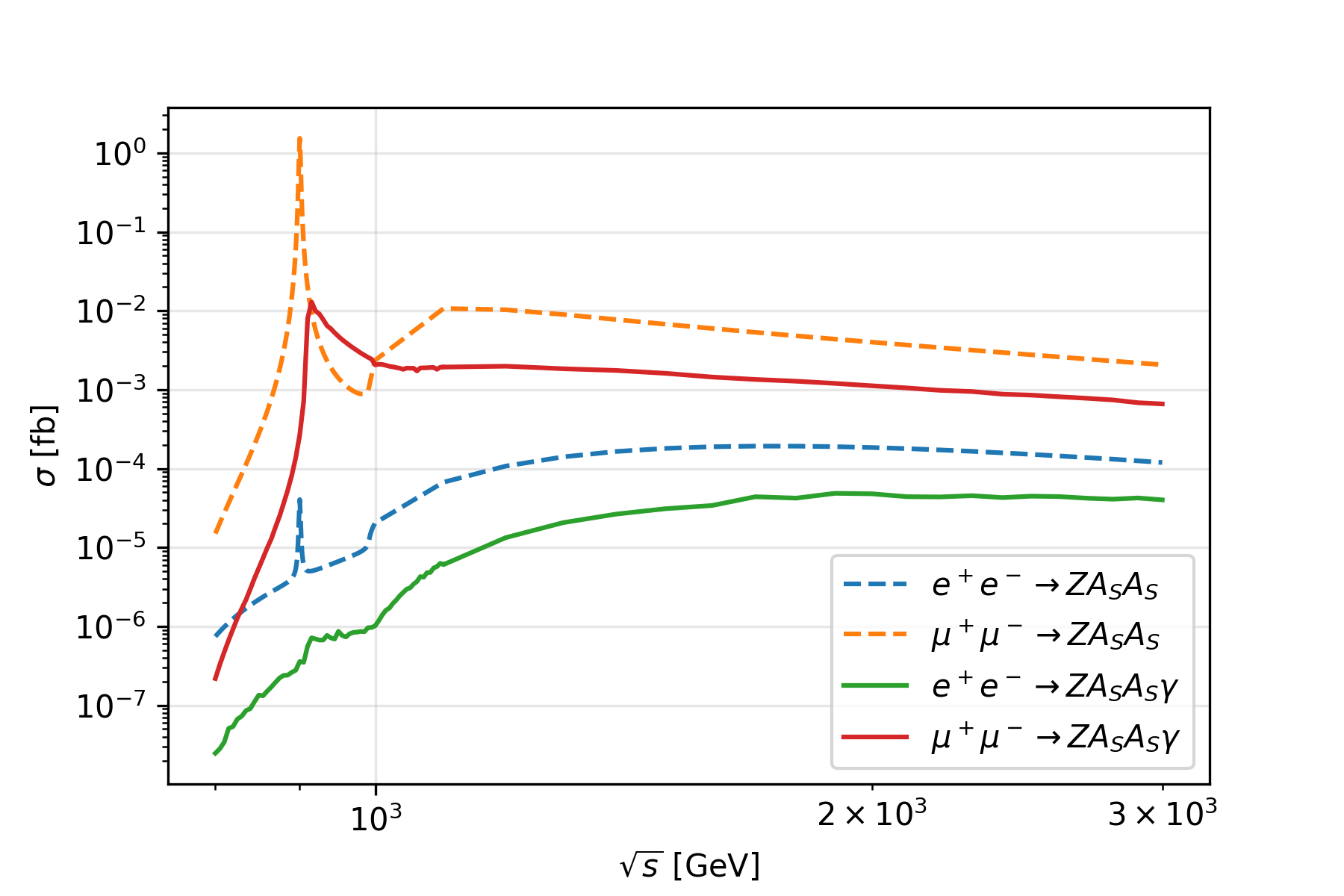}
    \caption{Production cross section for varying center of mass energy $\sqrt{s}$ at electron (blue and green lines) and muon (orange and red lines) collider. Shown are the results for production of two DM particles $A_S$ (left) without (dotted lines) and with (solid lines) a photon $\gamma$ in the final state and the same with an additional $Z$ boson in the final state (right).}
    \label{fig:emu} 
\end{figure}

In this section we look into the production cross sections of different final states including DM particles at the proposed future electron and at muon colliders under variation of the center of mass energy $\sqrt{s}$, for BP1. At these kind of colliders the heavy scalar $h_3$ could be produced directly from lepton and anti-lepton and then decay into two DM particles $A_S$. Possible final states include either only $A_S A_S$ or additional $Z$ bosons or photons $\gamma$. The results can be seen in figure \ref{fig:emu}.

On the left the processes $e^+e^-/\mu^+\mu^- \rightarrow A_S A_S$ are shown without and with a photon in the final state (initial state radiation) are shown. Since the DM particles cannot be detected, only the photon and missing energy would be measured.

On the right the processes $e^+e^-/\mu^+\mu^- \rightarrow Z A_S A_S$ are shown again without and with a photon in the final state. In this case the $Z$ boson and the photon and missing energy would be detected.

In both plots a peak around $\sqrt{s}\approx 900 \, \text{GeV} \approx m_{h_3}$ can be seen, as resonant production of one $h_3$ is possible, which would then decay into two $A_S$ in the final state. Furthermore in both plots the cross sections for the muon collider are higher than for the electron collider, which one would expect since the muon has stronger Yukawa couplings to the Higgs doublets.

For all processes with photons in the final state the following cut on the photon energy was employed: $E_\gamma > 10\,\text{GeV}$ and for the photon angle: $\theta>7^\circ$ in order to avoid divergences~\cite{Kalinowski:2020lhp}.

\subsubsection{HL-LHC}
For the analysis at HL-LHC, the main production channels for the heavy scalar $h_3$ are gluon gluon fusion (GGF) and vector boson fusion (VBF). The $h_3$ could then decay into two DM particles $A_S$. For both processes the significance $\mathcal{S} = \sqrt{2 \times \left[ (s+b){\rm ln}(1+\frac{s}{b})-s\right]}$ of the signal $s$ over the total SM background $b$ is calculated while taking into account some cuts which can be found in ref.~\cite{dutta2023dark}. The results are for BP1 are:
\begin{align}
    \text{GGF: } \mathcal{S} &= 1.356 \, \sigma \nonumber \\
    \text{VBF: } \mathcal{S} &= 0.007 \, \sigma , \nonumber
\end{align}
which is very low. However in another benchmark or by using machine learning techniques these results could be improved.

\section{Conclusions}
In this work we have investigated the DM phenomenology in a 2HDMS, where the DM candidate is the pseudo-scalar component of the singlet. We have found a benchmark point which is allowed under theoretical and experimental constraints and which also accommodates an excess found at CMS and LEP, which can be interpreted as a scalar Higgs particle with a mass of $95\,\text{GeV}$.

Furthermore we have looked into the production prospects at future lepton colliders and found potentially promising results. These could be further improved when taking beam polarization into account, as well as doing broader parameter scans and looking into more benchmarks. Future studies on this are in progress. The significance at HL-LHC however is rather low in the studied benchmark. Comprehensive parameter scans are planned for the future.

This work is a short summary of the study done in ref.~\cite{dutta2023dark}. Further information and explanations can be found there.

\section*{Acknowledgements}
JZ, JD, JL and GMP acknowledge support by the Deutsche Forschungsgemeinschaft (DFG, German Research Foundation) under Germany's Excellence Strategy EXC 2121 "Quantum Universe"- 390833306.  JD acknowledges support from the  HEP  Dodge Family Endowment Fellowship at the Homer L.Dodge Department of Physics $\&$ Astronomy at the University of Oklahoma.  
\bibliographystyle{JHEP}
\bibliography{ref}
\end{document}